\documentclass[useAMS,usenatbib]{mn2e}
\usepackage{times}   
\usepackage{graphicx}   
\usepackage{amsmath}   
\usepackage{natbib}   
\usepackage{color,ulem}  
\bibpunct{(}{)}{;}{a}{}{,}   

\title[Footprints in the sand]{Footprints in the sand: What can globular clusters tell us about NGC\,4753 past?}
\author[J. P. Caso et al.]{Juan Pablo Caso$^{~1,2}$\thanks{E-mail:
jpcaso@fcaglp.unlp.edu.ar (JPC), lbassino@fcaglp.unlp.edu.ar (LPB), matiasgomez@unab.cl (MG)} and Lilia P. 
Bassino$^{~1,2}$ and Mat\'ias G\'omez$^{~3}$\\
$^{1}$Grupo de Investigaci\'on CGGE, Facultad de Ciencias Astron\'omicas y Geof\'isicas de la Universidad Nacional de La Plata,    
and \\ Instituto de Astrof\'isica de La Plata (CCT La Plata -- CONICET, UNLP), Paseo del Bosque S/N,  
B1900FWA La Plata, Argentina\\   
$^{2}$Consejo Nacional de Investigaciones Cient\'ificas y T\'ecnicas, Rivadavia 1917, C1033AAJ  
Ciudad Aut\'onoma de Buenos Aires, Argentina\\   
$^{3}$Departamento de Ciencias Fisicas, Facultad de Ciencias Exactas, Universidad Andres Bello, Republica 252, Santiago, Chile.}

\date{Released 2002 Xxxxx XX}

\pagerange{\pageref{firstpage}--\pageref{lastpage}} \pubyear{2002}

\begin{document}

\label{firstpage}

\maketitle

\begin{abstract}
NGC\,4753 is a bright ($M_V\approx −22.3$) lenticular galaxy. It is a very 
interesting target to test different theories of formation of lenticular 
galaxies, due to its low-density environment and complex structure. We 
perform the first comprehensive study of NGC4753 globular cluster system (GCS), 
using Gemini/GMOS and CTIO/MosaicII images. Our results indicate a rather 
poor GCS of $\approx1000$ members. Its azimuthal distribution follows the 
shape of the galaxy bulge. The GC colour distribution is peculiar, presenting 
an intermediate subpopulation in addition to blue and red ones.
This intermediate subgroup can be explained by a single stellar population 
with an age of $1.5−3$\,Gyr and $0.5-1\,Z_{\odot}$. The GC specific frequency 
$S_N = 1.3\pm0.15$ is surprisingly low for a galaxy of its class. The GC 
luminosity function (GCLF) is also peculiar, with an excess of bright GCs 
compared to the expected gaussian distribution. The underlying galaxy body 
has significant substructure, with remnants of spiral arms, dust filaments, 
and isophote twisting. This, and the fact that NGC4753 hosted two type Ia 
SNe, support the possibility that the intermediate GC subpopulation may have 
originated during a recent merger, $1−3$\,Gyr ago.
\end{abstract}


\section{Introduction}
Despite that the fraction of lenticular galaxies (S0s) seems
to increase with the number of neighbours \citep{wil12b},
the existence of S0s in low density environments poses  
questions about their origin. 
The physical processes involved in the evolution of lenticular 
galaxies (S0s) have been largely discussed in the literature
\citep[e.g.][]{spi51,gun72,vbe09a}. Ram pressure stripping 
\citep[e.g.][]{mor06} and galaxy harrassment \citep{moo96} 
are currently considered as plausible explanations for the 
origin of S0s in high density environments. However, these 
processes are not expected to play a main role in low density 
environments. Alternative explanations are disk gas removal 
by an active galactic nucleus, or an inefficient inflow of 
external gas \citep[e.g.][and references therein]{jia09,vbe09b}.
However, cases  like NGC\,404 \citep{del04,thi10} and NGC\,4460 
\citep{moi10} seem to present a steady, long term gas accretion,
with current star formation, implying a complex evolutionary 
history.

NGC\,4753 is an early-type galaxy (ETG) with prominent and twisted dust 
lanes. It is considered as the brightest galaxy of one of the MK-groups 
\citep{kar13} that populate the Virgo Southern Extension, which is 
assigned 23 members. Alternatively, \citet{gar93} has included NGC\,4753 
in the one dozen-members' group LGG\,315 (or the 'NGC\,4643 group' 
according to NED, NASA/IPAC Extragalactic Database).
Originally classified as a peculiar lenticular galaxy
\citep{san61a}, it is now assigned an irregular de Vaucouleurs type I0 
(e.g. NED). 
This latter classification points to a merger or a galaxy distorted 
by gravitational interaction with a close neighbor, though the existence of a 
neighbor is not evident. The apparently 
high type Ia Supernovae (SNIa) rate exhibited by this galaxy (NGC\,4753 hosted SN\,1965I 
and SN\,1983G, \citealt{cia71} and \citealt{kos83,but85,mue94}, respectively) may 
be an indication of an intermediate-age population, 
a stellar population that may have formed in a merger a few Gyrs ago. 
However, this latter statement seems contradictory to the study performed 
by \citet{nav01}, who concluded that excepting strongly interacting 
systems, the SN production in isolated galaxies, galaxy pairs and groups 
does not depend on the environment. Or has NGC\,4753 possibly suffered 
such a strong event?, we will come back to this issue in the Discussion.

Regarding the dust content, \citet{ste92} proposed a disk model that is 
fitted to the dust distribution, so that the intrincated lanes can be 
understood as the result of a disk that has been twisted deeply by 
differential precession. In this way, the original S0 turns into a peculiar 
galaxy after an accretion event. Moreover, \citet{dew99} have estimated 
photometrically the mass of cold dust with optical and far-infrared (FIR) 
data, and showed that it is about a factor 10 higher than that estimated 
by mass-loss from only red giant stars. More recently, \citet{fin12} 
studied the relation between dust and ionized gas in ETGs with dust lanes, 
including NGC\,4753 in their sample. In order to explain the origin of the 
interstellar medium (ISM) detected in these ETGs, in addition to the 
internally-produced ISM, they stressed on the plausible capture of a gas-rich 
spiral or dwarf, or the collision between two gas-rich dwarfs with similar 
masses. Thus, we are again led to an accretion or merger event.

With the aim of disentangling the evolutionary story of NGC\,4753 and getting 
a new and precise distance determination,  we present a deep photometric 
study of its globular cluster (GC) system. It is well known that old 
globular cluster systems (GCS), i.e. with ages greater than 10\,Gyr, carry 
a historical record of the star formation events and, more generally, of 
the evolution of their host galaxies 
\citep[e.g.][and references therein]{bro06,ton13,kru14}. The formation of 
massive star clusters happens during major star-bursts. Thus, the GC colour 
distribution can provide evidence of star-burst episodes or merger events, 
which can be associated with the presence of an intermediate-age population 
\citep[e.g.][]{ric12a,cas13b}. For instance, \citet{sal15} have indicated 
that low values of the specific frequency $S_N$ \citep{har81}
or $T_N$, being $T_N$ the 
number of GCs normalized by the stellar mass of the host galaxy \citep{zep93},
are due to the paucity of red ({\it bona fide} high metal content) GCs. 
In the current scheme of GCs origin, blue ({\it bona fide} low metal 
content) GCs are formed at high-redshift, during the massive star 
formation episodes that follow the merging of the first building blocks 
of proto-galaxies. Subsequent merger events are responsible for the 
formation of red GCs, accompanied by another burst of few blue ones 
\citep{mur10,li14}. Alternatively, \citet{ton13} presents a hierarchical 
clustering model where the metal-rich subpopulation is composed of globular
clusters formed in the galaxy main progenitor around redshift $z\approx 2$, 
and the metal-poor subpopulation is composed of clusters accreted from 
satellites, and formed at redshifts $z\approx 3-4$. In this context, the 
study of GCSs will help to understand the origin and evolution of ETGs in 
low density environments.

Distance determinations for NGC\,4753 range between 20 and 24\,Mpc, including 
supernova light-curve analysis \citep{par00,rei05}, Tully-Fisher relation 
\citep{the07}, and surface brightness fluctuations \citep{ton01}. 
We will adopt a mean distance of $\approx 23.6$\,Mpc \citep{ton01} in the 
rest of this paper, until we can compare it with the result obtained from 
the GC luminosity function (GCLF). There are no previous distance measurements 
using the turn-over of the GCLF. The information on NGC4753's GCs is rather 
non-existent. \citet{zar15} include NGC\,4753 in their sample of 97 ETGs for which the 
number of GCs as well as radial profiles are estimated using Spitzer S4G data. 
However, as the authors indicate, these are rather crude estimates and focused
on the central region.

\medskip
This paper is organized as follows. The Observations are presented in 
Section\,2, the Results are shown in Section\,3, and the Section\,4 is
devoted to the Discussion.

\begin{figure}    
\includegraphics[width=83mm]{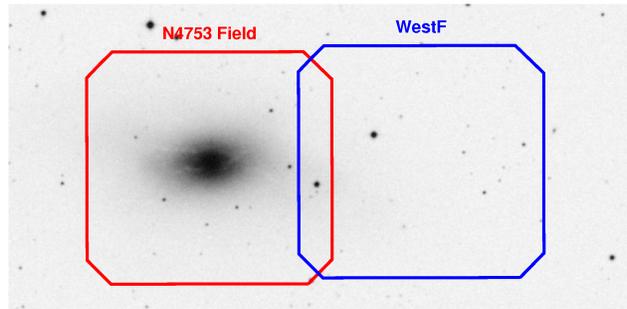}    
\caption{Positions of the two GMOS-S fields from our programme, 
superimposed to an $R$ image from the Palomar Observatory Sky
Survey. The image size is $15' \times 7.5'$. North is up, East 
to the left.}    
\label{fields}    
\end{figure}

\section{Observations and reduction}

\subsection{Observational data}
Our dataset consists of GMOS-Gemini-South images in
$g'$, $r'$, and $i'$ filters, observed during March/April
2014 (programme GS2014A-Q-67, PI: J.P. Caso). These images 
correspond to two fields, one of them centred on NGC\,4753
(hereafter, N4753F), and the other was located to the
 West (hereafter, WestF). Both fields were slightly 
overimposed, in order to determine possible differences in
the zero-magnitude (see Fig\,\ref{fields}). The exposure
times were $4\times 570s$ in $g'$, $4\times 270s$ in $r'$, 
and $4\times 360s$ in $i'$. Each set of exposures were
slightly dithered, in order to fill the gaps in the GMOS
field, and to efficiently remove cosmic rays and bad pixels.
The resulting seeing for the three filters was $0.6''-0.8''$.
The assumed absorptions come from NED \citep{sch11}.

The foreground contamination by Galactic stars was 
estimated from the observations of the William Herschel 
Deep Field (WHDF). This was necessary bevause the GCS
extended over the two science fields (see Section\,\ref{radsec}).
The reduced
observations of this field, part of the GMOS-N system 
verification observations (programme GN-2011B-SV-104), 
were downloaded from the Gemini Science Archive. This 
field had been observed in the same filters than our programme, 
with similar exposure times. According to its Galactic 
Latitude ($b= -61.7^{\circ}$), the foreground contamination 
would be similar to that of the N4753F ($b= 61.7^{\circ}$).
The reduction was perfomed in the usual way, using the
tasks from the GEMINI-GMOS package, within \textsc{iraf}.
It was necessary to subtract the {\it fringe} pattern in
the $i'$ images from our programme. The bias, flatfield, 
and {\it i' blank sky} images were obtained from the Gemini 
Science Archive.

This dataset is further enhanced by using images in 
Kron-Cousins $R$ filter, obtained 
with the MOSAIC\,II camera mounted at the 4-m Blanco 
telescope at the Cerro Tololo Inter-American Observatory 
(CTIO, Chile). The observations were carried out on
March 25--26, 2004. The observation was split 
in four exposures of 720\,s each. During these nights, one 
of the MOSAIC\,II CCDs was not working. To overcome this,
individual images were taken with an appropiate 
dither to fill in the missing chip. Thus, the final image 
presents a field-of-view of $36 \times 45$\,
arcmin$^2$, i.e., $\approx 247\times 309\,{\rm kpc}^2$ at 
the assumed distance to NGC\,4753.

\subsection{Photometry and point source selection}
First, the extended galaxy light was substracted from the
GMOS observations of the N4753F and WestF, applying a 
ring median filter \citep{sec95} with an inner radius of 
5\,arcsec (35\,px), and an outer radius of 6\,arcsec (41\,px). 
This procedure facilitates the point source detection, and is commonly used 
in GCs studies \citep[e.g.][]{dir03a,bas06a,bas06b,cas13a,cas14}.
This procedure was not necessary in the case of the WHDF.

The software SE\textsc{xtractor} \citep{ber96} was applied to 
the $i'$ image (because it presented the higher 
signal-to-noise ratio), in order to obtain an initial 
point source catalogue. The effective radii of clasical GCs
is usually smaller than 10\,pc \citep[e.g.][]{bru12}.
Considering that, at the assumed distance, $1''$ corresponds
to $\approx 110$\,pc, the NGC\,4753 GCs are seen as point 
sources on our images. Then,
we used the SE\textsc{xtractor} parameter CLASS$\_$STAR to
eliminate the extended sources from our catalogue.

The photometry was performed with the DAOPHOT package \citep{ste87}
within \textsc{iraf}. A second-order variable PSF was generated
for each filter, employing a sample of bright stars, well distributed
over the field. The final point source selection was made with the
$\chi^2$ and sharpness parameters, from the ALLSTAR task.

\subsection{Photometric calibration}
During our programme, a well populated standard 
star field, from the list of \citet{smi02} 
was observed during the same night than the N4753F. We obtained
the standard stars aperture photometry for several aperture radii
with task \textsc{phot}. We considered these measurements to derive the 
aperture correction by using a growth curve. After this correction, 
we fit transformation equations of the form:

\begin{equation}
m_{std} = ZP + m_{inst} - K_{CP} \times (X-1) + CT \times (m_1 - m_2)_{std} \nonumber
\end{equation}

\noindent where $m_{std}$ and $m_{inst}$ are the standard and 
instrumental magnitudes, respectively, and $ZP$ are the 
photometric zero points. $K_{CP}$ is the mean atmospheric extintion 
at Cerro Pach\'on, obtained from the Gemini Observatory Web 
Page\footnote{http://www.gemini.edu/sciops/instruments/gmos/calibration}, 
and $X$ the airmass. $CT$ is the coefficient of the color term,
 and $(m_1 - m_2)_{std}$ its associated calibrated colour, $(g'-r')$ 
for $g'$ and $r'$, and $(r'-i')$ for $i'$. The fitted zero points 
and colour terms were $ZP^{g'}=28.185\pm0.013$, 
$ZP^{r'}=28.314\pm0.009$ and $ZP^{i'}=27.915\pm0.010$, and
$CT_{g'}=0.106\pm0.018$, $CT_{r'}=0.070\pm0.012$ and $CT_{i'}=0.039\pm0.023$,
respectively. The calibration equations for the WHDF were
obtained from \citet{fai11}.

Afterwards, we applied the galactic extinction corrections from 
\citet{sch11} to the calibrated magnitudes. Finally, from the objects 
in common between the N4753F and WestF we found small zero-point 
differences, $\Delta_{g'}=-0.05$, $\Delta_{r'}=0.04$ and $\Delta_{i'}=0.02$. 
These offsets were applied to the WestF final catalogue, in order to 
refer its photometry to the N4753F one.

In the case of the MOSAIC\,II observations, 4 to 5 fields were 
observed in Kron-Cousins $R$ and Washington $C$ filters,
containing about 10 standard stars from the 
list of \citet{gei96b}. The observations spanned a large range in airmass,
typically from 1.0 to 1.9. The coefficients fitted for each night were 
indistinguishable within the uncertainties. Considering that, it was possible 
to calculate a single set of transformation equations, which resulted

\begin{equation}
  C =  c - 0.059\pm0.004 - 0.418\pm0.010 \times X_C + 0.111\pm0.005 \times (C-T_1) \nonumber
\end{equation}
\begin{equation}
  T_1 = r + 0.628\pm0.005 - 0.14\pm0.002 \times X_R + 0.019\pm0.002 \times (C-T_1) \nonumber
\end{equation}
where ${r}$ and $c$ are the instrumental magnitudes, and 
$X_\mathrm{R}$ and $X_\mathrm{C}$ are the airmasses in $R$ and $C$ filters,
respectively.

\subsection{Completeness analysis}
\begin{figure}    
\includegraphics[width=35mm,angle=270]{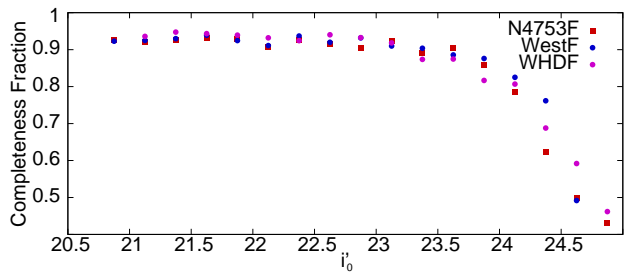}    
\caption{Completeness curves for the three fields (our science fields 
plus the WHDF), as function of $i'_0$ magnitude. The bin width is 0.25\,mag.}    
\label{comp}    
\end{figure}

In order to estimate the photometric completeness for the three fields
(our science fields plus the WHDF),
we added 250 artificial stars in the images of the three filters.
These stars were equally distributed over the entire GMOS-S fields, 
they presented colours in the expected ranges for GCs, and $i'> 20.75$. 
This process was repeated 40 times, achieving a sample of 10\,000
artificial stars.
Their photometry was carried out in the same way as the science fields.
The completeness curve, after the definite point source catalogue was 
generated, is shown for both fields in Figure\,\ref{comp}. 
The completeness functions are very similar, achieving the $80\%$ at 
$i'\approx 24.1$. This value was considered as the faint magnitude 
limit in the following analysis.

The comparison between the initial brightness of artificial stars and 
their subsequent photometry proved that the application of the median filter
does not modify the point source photometry.

\section{Results}

\subsection{GC candidates selection from colours}
\begin{figure}    
\includegraphics[width=80mm]{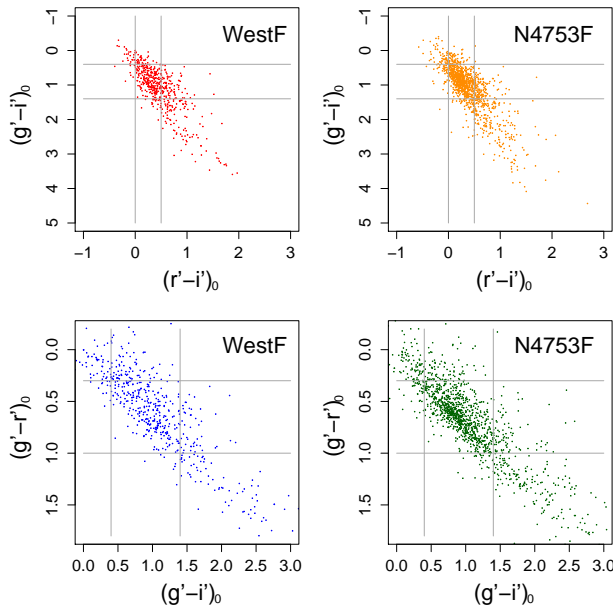}    
\caption{Colour-colour diagrams for both science fields. Solid lines indicate
the colour ranges of GC candidates (see Section\,3.1).}   
\label{dcc}    
\end{figure}

Figure\,\ref{dcc} shows $(g'-i')_0$ vs. $(r'-i')_0$ and $(g'-r')_0$
vs. $(g'-i')_0$ colour-colour diagrams (CCD) for point sources in
both fields. Old GCs present narrow colour ranges, which can be helpful to
distinguish them from contamination in the CCDs \citep[e.g.][]{fai11,esc15}.
It is particularly evident in the N4753F, where it is expected
to find the majority of the GCs belonging to NGC\,4753 system. We
selected as GC candidates the point sources ranging
$0.4 < (g'-i')_0 < 1.4$, $0.3 < (g'-r')_0 < 1$ and $0 < (r'-i')_0 < 0.5$.
These values are similar to those chosen by \citet{fai11} and \citet{esc15}.
The number of point sources brighter than $i'=24.1$ (our 80\% completeness limit, 
see above) that fulfill the
colour criteria are 437 in the N4753F, 117 in the WestF, and
32 in the WHDF.

\begin{figure}    
\includegraphics[width=80mm]{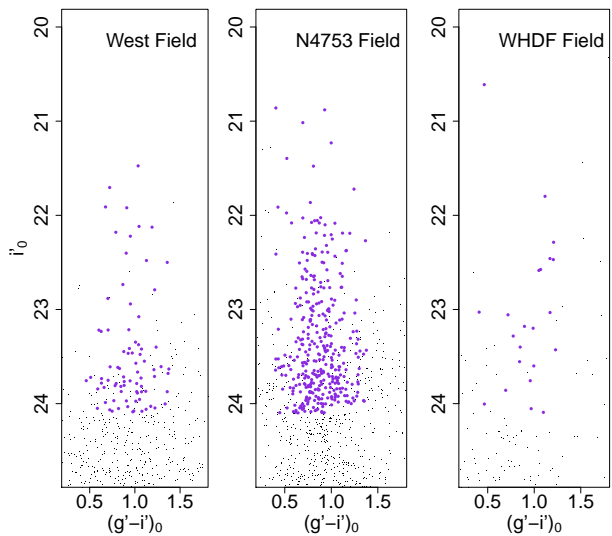}    
\caption{Colour-magnitude diagrams for the three fields. Black dots represent
point sources, while blueviolet filled circles highlight those  objects that 
fulfill the colours and magnitude criteria applied to select GC candidates.}    
\label{dcm}    
\end{figure}

In Figure\,\ref{dcm} we show the $i'_0$ vs. $(g'-i')_0$ colour-magnitude
diagrams for the three fields. The black dots represent the point sources, 
after the $\chi^2$ and sharpness selection. The blueviolet 
filled circles highlight those objects that fulfill the colours and magnitude 
criteria indicated before.
The CMD in the middle panel, corresponding to N4753F, shows that GC
candidates span a rather large range, with a large fraction of blue GCs. 
Comparing the CMDs from the WestF and WHDF, it can be seen that some GC 
candidates are still present in the former one. If we assume that the 
brightness limit between ultra-compact dwarfs (UCDs) and `classical' GCs
is $M_V \approx -10.5 - -11$ \citep[e.g.][]{mie06,hil09b,cas13a,cas14},
at NGC\,4753 distance implies $i'_0 \approx 20.9 - 20.4$. Hence, 
NGC\,4753 does not seem to present any candidate to UCD.

\subsection{Colour distribution}
\label{dcolsec}

\begin{figure}    
\includegraphics[width=85mm]{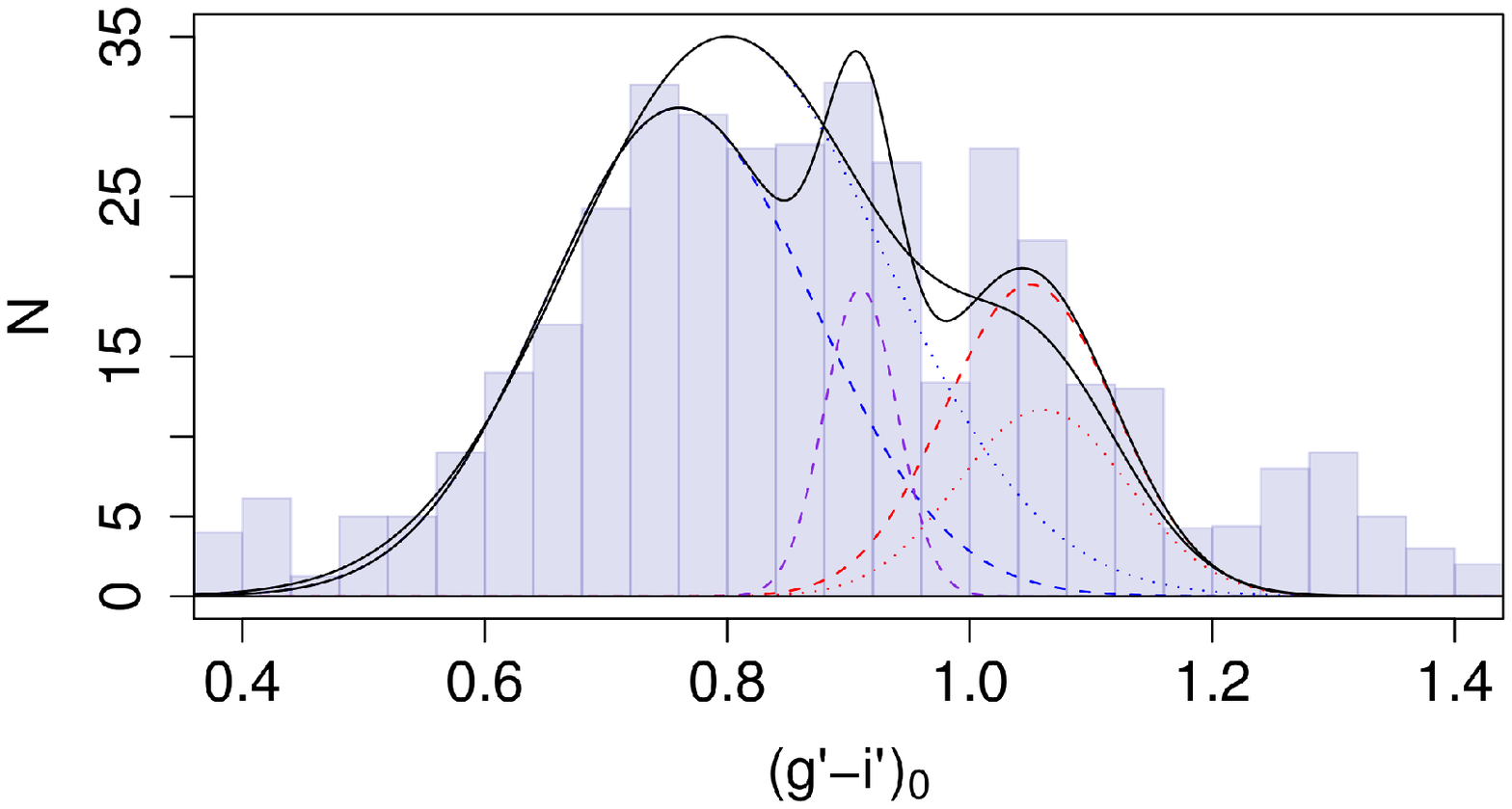}\\    
\includegraphics[width=85mm]{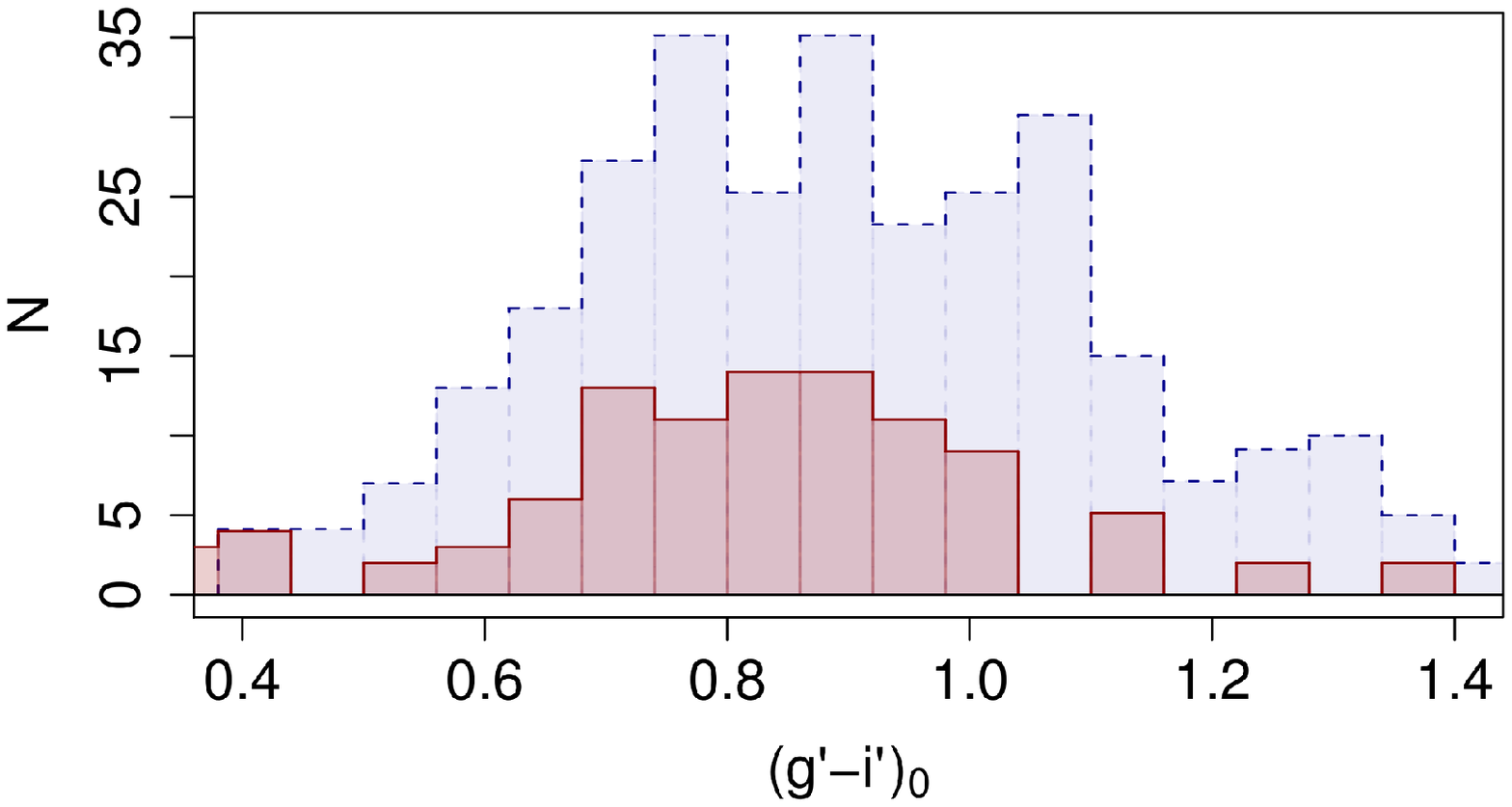}\\    
\caption{{\bf Upper panel:} Colour distribution for the GC candidates. 
The dotted lines indicate the Gaussian fits to the {\it bona fide} 
metal-poor population and metal-rich population assuming 
a bimodal distribution. For three components, the Gaussian fits are 
shown with dashed lines. The solid lines represent the sum of Gaussians
for each scenario. {\bf Lower panel:} Colour distribution for 
GCs candidates in the magnitude range $24.1 > i'_0 > 23$ (dashed line 
histogram) and $i'_0 < 23$ (solid line histogram).}    
\label{dcol}    
\end{figure}

The Figure\,\ref{dcol} shows the colour distribution for the GC 
candidates with $i'_0< 24.1$, applying a binwidth of 0.04\,mag. The colour 
distribution seems to deviates from the typical bimodal pattern, frequently
seen  in early-type galaxies \citep[e.g.][and references there in]{for14,bro14}. 
We fit means and dispersions for bimodal and trimodal distributions,
using the Gaussian Mixture Method \citep[{\textsc GMM},][]{mur10}. The
results are listed in Table\,\ref{fit}, where we also include the 
fractions of each subpopulation $f$ as well as the DD parameter, that 
indicates whether an specific multimodal distribution provides a realistic 
fit (where the null hypothesis is accepted for DD$>2$). For the 
three components distribution, the single Gaussians are plotted with dashed 
lines. Doted lines represent the Gaussians for the bimodal distribution. 

If we consider all GC candidates together (first and second rows in 
Table\,\ref{fit}), we can see that for the trimodal distribution, the bluer 
and redder Gaussian means resemble those of blue and red GC (metal-poor and 
metal-rich) subpopulations, respectively \citep[e.g.][]{fai11,sal15}. However, 
there is evidence of a third population, presenting intermediate colours. 
This may reflect a complex evolutionary history, similar to other early-type 
galaxies in low-density environments \citep[e.g.][]{cas13b,esc15}. The
intermediate GC subpopulation presents mean $(g'-i')_0 \approx 0.9$ in all
cases (Table\,\ref{fit}). We compare this colour with the theoretical models 
of single stellar populations (SSPs) by \citet{bre12}, using their web-based 
tool \footnote{http://stev.oapd.inaf.it/cgi-bin/cmd}. If we consider SSPs
with metallicities $0.5\,Z_{\odot}$ and $Z_{\odot}$, a \citet{cha01} lognormal
IMF, and absorption $A_V=0.09$ (NED), the mean colour for the intermediate
subpopulation corresponds to a population of $1.5-3$\,Gyr. This age is
consistent with the existence of SNIa in the galaxy population. The
derived ages shoul be interpreted as raw estimations. The selection
of these particular metallicities, as well as the IMF, might introduce
systematic errors.

Another interesting issue is that if we consider the bimodal distribution,
the mean value, dispersion and fraction of blue GCs are in agreement, within 
uncertainties, with the equivalent values of blue GCs from the trimodal 
distribution. However, the fraction of red GCs in the bimodal case, seems to 
include approximately the intermediate plus the red subpopulations of the 
trimodal case. That is, the blue subpopulation remains almost unchanged. 
This might be the reason for the overestimation of the red population in 
Figure\,\ref{dcol}.

In order to look for radial gradients, we separate the sample in two
radial regimes. We assume an elliptical projected spatial distribution, 
with PA and ellipticity equal to $91^o$ and $0.45$ (see Section\,\ref{radsec}). 
The adopted radial galactocentric limit is taken as $R_g= 2.5$\,arcmin, 
which results in two samples with similar number of GC candidates.
The results for both radial ranges are depicted in the four following 
rows of Table\,\ref{fit}. In both cases, the $DD$ values from {\textsc GMM} 
indicate that a trimodal profile could represent the colour distribution.
The means and dispersions are similar, but the intermediate subpopulation
dominates the distribution when we move towards larger galactocentric
distances, at the expense of the blue subpopulation.

The comparison between faint and bright GCs also reveals different behaviours
in the colour distribution. The selection of $i'_0= 23$ as the adopted 
magnitude limit to separate bright and faint GCs is based on the GCs CMD 
(Fig.\,\ref{dcm}), and luminosity function (GCLF) (see Section\,\ref{lfsec}. 
Despite of the similarity in means and dispersions,
the weight of the different populations in the colour distribution differs
significantly. The faint GCs seem to present a trimodal distribution,
as well as the entire sample. The presence of the red population
in the bright sample seems marginal (lower panel of Fig\,\ref{dcol}).

\begin{table*}
\begin{center}
\caption{Parameters of the {\textsc GMM} fitting to the colour distribution, considering bimodal and trimodal cases. $\mu_i$, $\sigma_i$ and $f_i$
correspond to the means, dispersions and fractions for each Gaussian component. The parameter DD indicates whether an specific multimodal distribution 
provides a realistic fit (where the null hypothesis is accepted for DD$>2$).}
\label{fit}
\resizebox{\textwidth}{!}{   
\begin{tabular}{@{}lcccccccccc@{}}   
\hline   
\multicolumn{1}{@{}c}{}&\multicolumn{1}{c}{$\mu_{1}$}&\multicolumn{1}{c}{$\sigma_{1}$}&\multicolumn{1}{c}{$f_{1}$}&
\multicolumn{1}{c}{$\mu_{2}$}&\multicolumn{1}{c}{$\sigma_{2}$}&\multicolumn{1}{c}{$f_{2}$}&\multicolumn{1}{c}{$\mu_{3}$}&
\multicolumn{1}{c}{$\sigma_{3}$}&\multicolumn{1}{c}{$f_{3}$}&\multicolumn{1}{c}{DD}\\   
\hline   
\multicolumn{11}{c}{All GC candidates}\\   
\hline    
Trimodal  & $0.76\pm0.05$&$0.11\pm0.02$&$0.63\pm0.13$&$0.91\pm0.04$&$0.03\pm0.02$&$0.11\pm0.05$&$1.05\pm0.02$&$0.07\pm0.01$&$0.25\pm0.06$&$2.04$\\
Bimodal  & $0.80\pm0.04$&$0.13\pm0.02$&$0.81\pm0.15$&$1.06\pm0.05$&$0.07\pm0.02$&$0.19\pm0.15$&&&&$2.50$\\
\hline   
\multicolumn{11}{c}{$R_g < 2.5'$}\\   
\hline    
Trimodal  & $0.76\pm0.05$&$0.11\pm0.02$&$0.76\pm0.07$&$0.91\pm0.03$&$0.03\pm0.01$&$0.13\pm0.06$&$1.04\pm0.02$&$0.05\pm0.01$&$0.11\pm0.02$&$2.15$\\
Bimodal  & $0.80\pm0.03$&$0.12\pm0.01$&$0.95\pm0.04$&$1.06\pm0.05$&$0.04\pm0.02$&$0.05\pm0.04$&&&&$2.85$\\
\hline    
\multicolumn{11}{c}{$R_g > 2.5'$}\\   
\hline    
Trimodal  & $0.69\pm0.05$&$0.09\pm0.03$&$0.33\pm0.10$&$0.89\pm0.04$&$0.06\pm0.03$&$0.31\pm0.12$&$1.06\pm0.03$&$0.07\pm0.01$&$0.36\pm0.09$&$2.60$\\
Bimodal  & $0.68\pm0.08$&$0.08\pm0.03$&$0.41\pm0.17$&$0.96\pm0.05$&$0.10\pm0.03$&$0.59\pm0.17$&&&&$2.55$\\
\hline    
\multicolumn{11}{c}{$i'_0 < 23$}\\   
\hline    
Trimodal  & $0.74\pm0.04$&$0.07\pm0.02$&$0.54\pm0.08$&$0.92\pm0.04$&$0.05\pm0.02$&$0.40\pm0.06$&$1.14\pm0.05$&$0.03\pm0.01$&$0.06\pm0.01$&$2.92$\\
Bimodal  & $0.69\pm0.06$&$0.15\pm0.04$&$0.44\pm0.14$&$0.87\pm0.13$&$0.06\pm0.01$&$0.56\pm0.14$&&&&$2.10$\\
\hline    
\multicolumn{11}{c}{$23 < i'_0 < 24.1$}\\   
\hline    
Trimodal  & $0.75\pm0.03$&$0.11\pm0.02$&$0.57\pm0.10$&$0.91\pm0.02$&$0.04\pm0.01$&$0.14\pm0.07$&$1.06\pm0.01$&$0.06\pm0.01$&$0.29\pm0.04$&$2.20$\\
Bimodal  & $0.80\pm0.03$&$0.13\pm0.02$&$0.77\pm0.09$&$1.06\pm0.03$&$0.06\pm0.02$&$0.23\pm0.09$&&&&$2.62$\\
\hline   
\end{tabular}  
}
\end{center}
\end{table*}

Thus, both bimodal and trimodal distributions seem to represent the 
colour distribution with different degree of accuracy. However, the dependences 
of the distribution with galactocentric radius and brightness point to a 
GCS built with three different components.

\subsection{Azimuthal and radial distributions}
\label{radsec}

\begin{figure}    
\includegraphics[width=85mm]{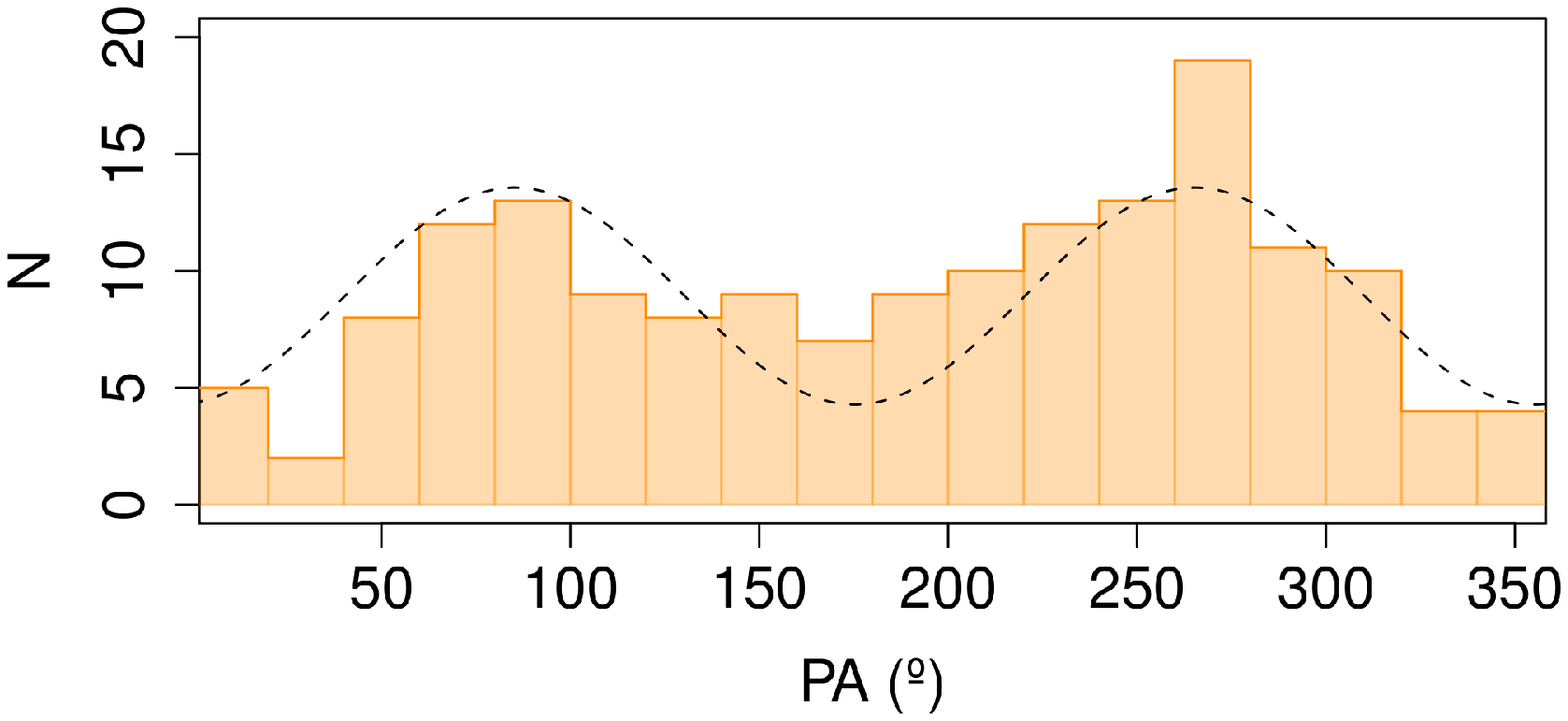}\\    
\includegraphics[width=85mm]{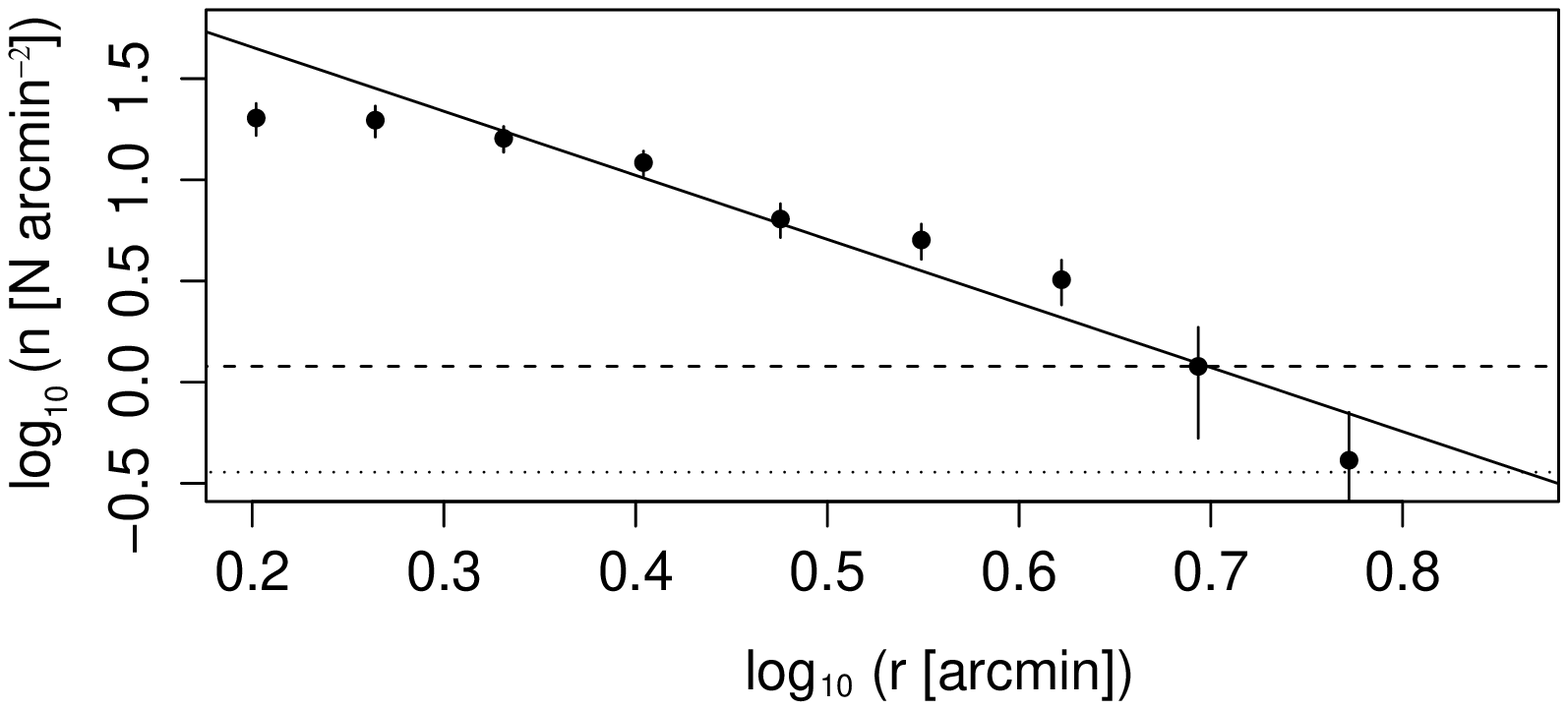}\\    
\caption{{\bf Upper panel:} azimuthal distribution for GC candidates
at less than 2.5\,arcmin from the galaxy centre. The GCs position 
angle (PA) is measured from North through East, and the bin width is 
$18^o$. {\bf Lower panel:} radial distribution for GC candidates. The
solid line represents the power-law fitted by least-squares. The dashed
line coresponds to the background level, while the dotted one indicates
the $30\%$ of the background level, which is the used to define the GCS 
extension.}
\label{dacim}    
\end{figure}

The upper panel of Figure\,\ref{dacim} shows the azimuthal distribution
for all the GC candidates at less than 2.5\,arcmin from the galaxy centre. 
This limit is chosen to achieve an homogeneous azimuthal coverage,
according to the field-of-view of the GMOS field. The position angle 
(PA) of the GCs projected distribution is measured from North through East, 
and the bin width is $20^o$. Considering that an elliptical GCS causes
a sinusoidal distribution in this diagram, we fit a cosine function
to the histogram. The fitted PA is $94\pm6^o$, which is in agreement
with the galaxy light PA (see Section\,\ref{galsec}). The amplitude
and the symmetry axis offset are $4.7\pm0.9$ and $8.9\pm0.7$, respectively.

The lower panel of Figure\,\ref{dacim} exhibits the logarithm of the 
background corrected radial distribution for the GC candidates. The 
population of the GCS is not large enough to discriminate between different
subpopulations.

Due to the elongated spatial distribution of the GCS, we considere
concetric ellipses instead of circles. The ellipses PA is the previously
fitted one, while the ellipticity was assumed as $\approx 0.45$, obtained
from the galaxy diffuse light (see Section\,\ref{galsec}). The horizontal
dashed line indicates the background level, while the dotted one is the
$30\%$ of it. We propose this
limit to define the GCS extension, which has been previously used in
similar studies \citep[e.g.][]{bas06a,cas13b}. 
The background corrected distribution can be fit by a power-law, with 
slope $-3.2\pm0.3$. According to this, the total extension of the GCS
is $\approx8$\,arcmin (i.e., $\approx 50$\,kpc). The inner bin of the 
radial distribution seems to underestimate the GCs density. This is 
probably due to a higher imcompleteness, caused by the high surface
brightness of the galaxy, plus its striking dust structure.

\begin{figure}    
\includegraphics[width=85mm]{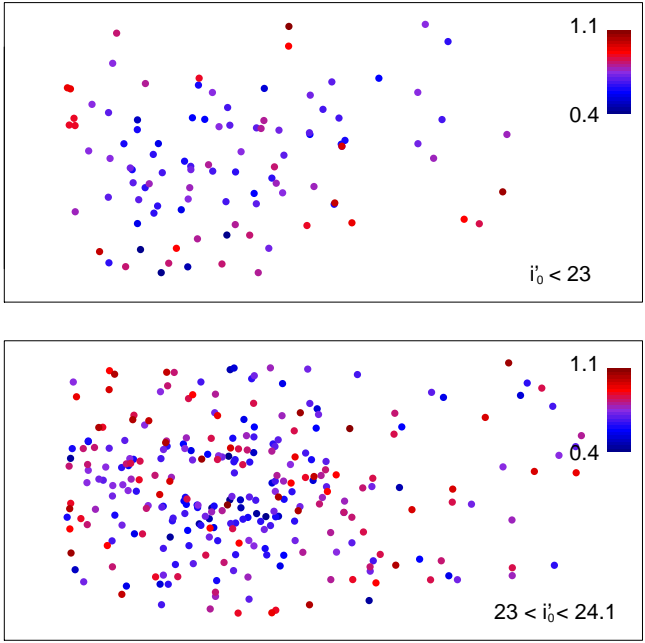}    
\caption{Projected spatial distribution for GC candidates, split in two
magnitude regimes. The centre of the galaxy is highlated with a cross. North 
is up, and East to the left. The field of view is $9.1\times5.5\,{\rm arcmin}^2$. 
The colour gradient spans the range $0.4< (g'-i')_0 <1.4$.}    
\label{espa}    
\end{figure}

The projected spatial distribution for GC candidates for two
different magnitude bins is shown in Figure\,\ref{espa}. The 
colour gradient in both panels spans the range $0.4< (g'-i')_0 <1.4$.
North is up, and East to the left. 
It seems that GC candidates brighter than $i'_0= 23$, mainly
blue candidates, present inhomogeneities in their spatial 
distribution. In particular,
there is a paucity of GCs to the North of NGC\,4753. 
The faint GC candidates present a much homogeneus spatial distribution,
although they follow the major axis of the galaxy
(approximately in the East-West direction). It can be seen GCs from
the three subpopulations at large projected distances from the the 
galaxy. This deviates from
the usual scenario, where the blue GCs are more spread, following 
the X-ray gas emission \citep[e.g.][]{for05,for12,esc15}, and the red 
ones are more concentrated, following the galaxy light
\citep[e.g.][]{bas06a}.

\subsection{Luminosity function and total number of GCs}
\label{lfsec}
\begin{figure}    
\includegraphics[width=85mm]{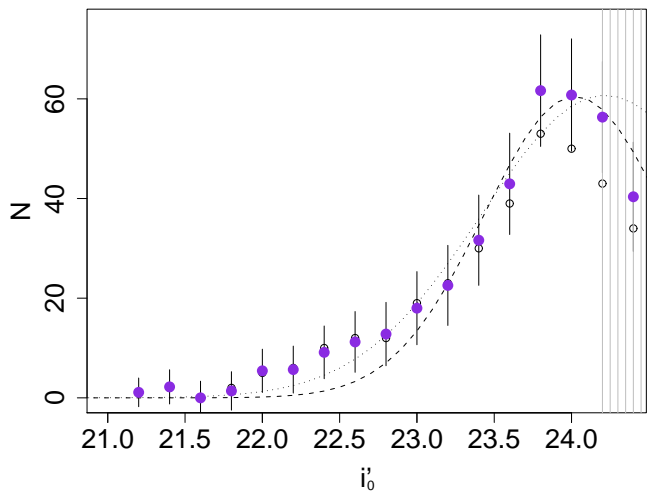}\\    
\caption{Luminosity function (GCLF) for GCs in the radial 
regime $1.25-3.6$\,arcmin (assuming an elliptical geometry).
Open circles represent the raw GCLF, while blueviolet
filled circles indicate the background and completeness 
corrected one. The errorbars assume Poisson 
uncertainties for science and background measurements, and
the binwidth is 0.2. The vertical lines indicate the 
luminosity range that has not beenconsidered in the GCLF 
fitting.}    
\label{lf}    
\end{figure}

In order to facilitate the calculus of the total number
of GCs, we obtain the GCLF for those
GCs located within two concentric ellipses with the same
PA and ellipticity, equal to $91^o$ and $0.45$, respectively
(see Section\,\ref{radsec}), and major semiaxes 1.25 and 
3.6\,arcmin. Figure\,\ref{lf} shows the raw GCLF
with open circles, and the background and completeness corrected 
one with blueviolet filled circles. The errorbars assume Poisson 
uncertainties for science and background measurements, and
the binwidth is 0.2. The vertical lines indicate the luminosity
range that has not been considered in the GCLF fitting, due the declining 
completeness (we selected the limit $i'_0=24.1$ in Section\,2.4).
It can be noticed a change in the GCLF
behaviour around $i'_0 \approx 23$, showing an excess of bright
GCs in comparison with faint ones.
We fitted a Guassian profile to the entire GCLF, which is
represented by the dotted curve. The 
resulting turn-over magnitude (TOM) and dispersion are
$i'_{0,TOM}=24.2\pm0.2\pm0.1$ and $\sigma=0.8\pm0.12\pm0.1$, 
where the second uncertainty comes from the
binning considered here. If we fit just the GCLF bins
with $i'>23$, we obtain a Gaussian profile slightly 
different, which is plotted with a dashed curve. Comparing
both profiles, it is clear that the excess of bright GCs
results in a higher dispersion for the fitted Gaussian, and
consequently the TOM moves slightly towards fainter values. The
TOM and dispersion in this latter case are 
$i'_{0,TOM}=24\pm0.08\pm0.1$ and $\sigma=0.59\pm0.07\pm0.1$,
respectively. In the following, we will assume that these 
parameters represent the GCLF for NGC\,4753.

Usually, old GC populations present a Gaussian GCLF, with a 
TOM in the $V$-band of $M_{TOM}\approx-7.4$
\citep[e.g.][]{ric03,jor07}. The GC candidates fainter than $i_0'=23$
present a mean colour $(g'-i')\approx0.9\pm0.03$, and then 
$g'_{0,TOM}\approx24.9\pm0.12$. Applying Equation\,2 from 
\citet{fai11}, the TOM in the $V$ filter is $V_{0,TOM}\approx24.55
\pm0.14$, which implies that the distance moduli and metric 
distance result $(m-M)\approx31.95\pm0.14$ and $\approx24.5\pm1.5$\,Mpc,
respectively. This distance is in agreement with previous measurements
from SBF \citep{ton01} and SNIa light-curves \citep{par00,rei05}.
GCLF dispersion seem to be lower than expected.

The numerical integration of the Gaussian profile results in
$445\pm 35$ GCs. We have to consider that the GCLF was 
obtained from the GC candidates in a specific radial regime,
that does not span the entire GCS extension. From the density
profile derived in Section\,\ref{radsec}, we calculate the
ratio of GCs in the radial regime $1.25-8$\,arcmin to
GCs in the radial regime $1.25-3.6$\,arcmin, $\approx1.26\pm0.07$.
Due to the incompleteness, we cannot derive the surface
density of GCs in the inner $1.25$\,arcmin. A common 
criterion in GC studies is to adopt for this region the same 
surface density as for the first bin \citep[e.g.][]{bas06b,cas13b},
and correct it for completeness. This can be done,
assuming that the radial distribution of GCs flattens 
in the inner region of a galaxy \citep[e.g.][]{els98}.
Hence, we obtain $1030\pm 120$ GCs. We have not considered yet 
the excess of 
bright GCs. We can do that subtracting to each GCLF bin the 
numerical integration of the Gaussian profile fitted to the
GCs fainter than $i'=23$. The sum of this excess result in
$\approx 40$ GCs. Then, the number of members of the 
NGC\,4753 GCS is $1070\pm 120$. Assuming the total $V$ magnitude
obtained by the Carnegie-Irvine Galaxy Survey \citep{ho11},
$V_{tot}= 9.66\pm0.11$, and the distance moduli previously 
derived, $m-M= 31.95\pm0.14$, the absolute
magnitude of the galaxy is $M_V=-22.3\pm0.2$. Then, the
specific frequency results $S_N=1.3\pm 0.15$.

\subsection{NGC\,4753 surface photometry}
\label{galsec}

NGC\,4753 presents several dust filaments crossing its centre,
mainly in the East-West direction, as well as faint plums in 
the outskirts of the galaxy. Since their presence could point 
to a distorted structure, we obtained NGC\,4753 surface 
brightness profiles in the Gemini GMOS-S $g'$, $r'$ and $i'$ 
exposures and the MOSAIC\,II $R$ image applying the task ELLIPSE, 
within {\textsc iraf}.

\begin{figure}    
\includegraphics[width=85mm]{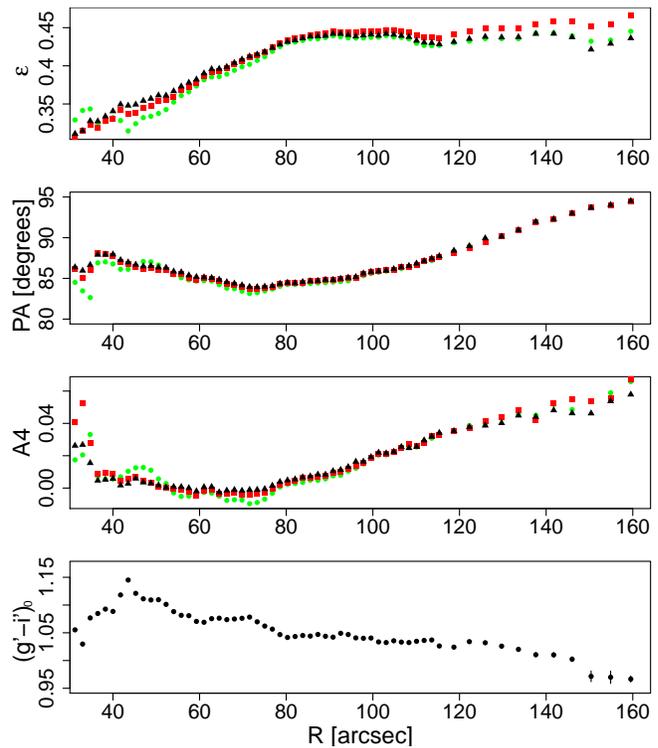}\\    
\caption{From top to bottom, ellipticity, position angle, harmonic A4
and $(g'-i')_0$ colour profile in the radial regime $30-160$\,arcsec,
obtained from our GMOS-S observations. In the three upper panels,
green circles, red squares and black triangles represent parameters 
from $g'$, $r'$ and $i'$ filters, respectively.}    
\label{param}    
\end{figure}

Figure\,\ref{param} shows the radial dependence of some important 
morphological parameters, derived for $g'$, $r'$ and $i'$ filters. 
No striking difference is seen among them in ellipticity, 
position angle or the higher harmonic A4 associated to diskyness/boxyness. 
The strong isophote twisting which is present in the inner 
part of the galaxy (at radii smaller than 40\,arcsec) is not shown 
since ellipses are not representative here. It is apparent that the 
variation in the PA at small radii is wider in $g'$ than $i’$, as expected 
from dusty substructures. The parameter A4 is mainly positive, which
is expected in disky isophotes, from rotationally supported galaxies.
The $(g'-i')_0$ colour profile presents a slight gradient, becoming
bluer towards the outskirts.

\begin{figure}    
\begin{flushright}
\includegraphics[width=78mm]{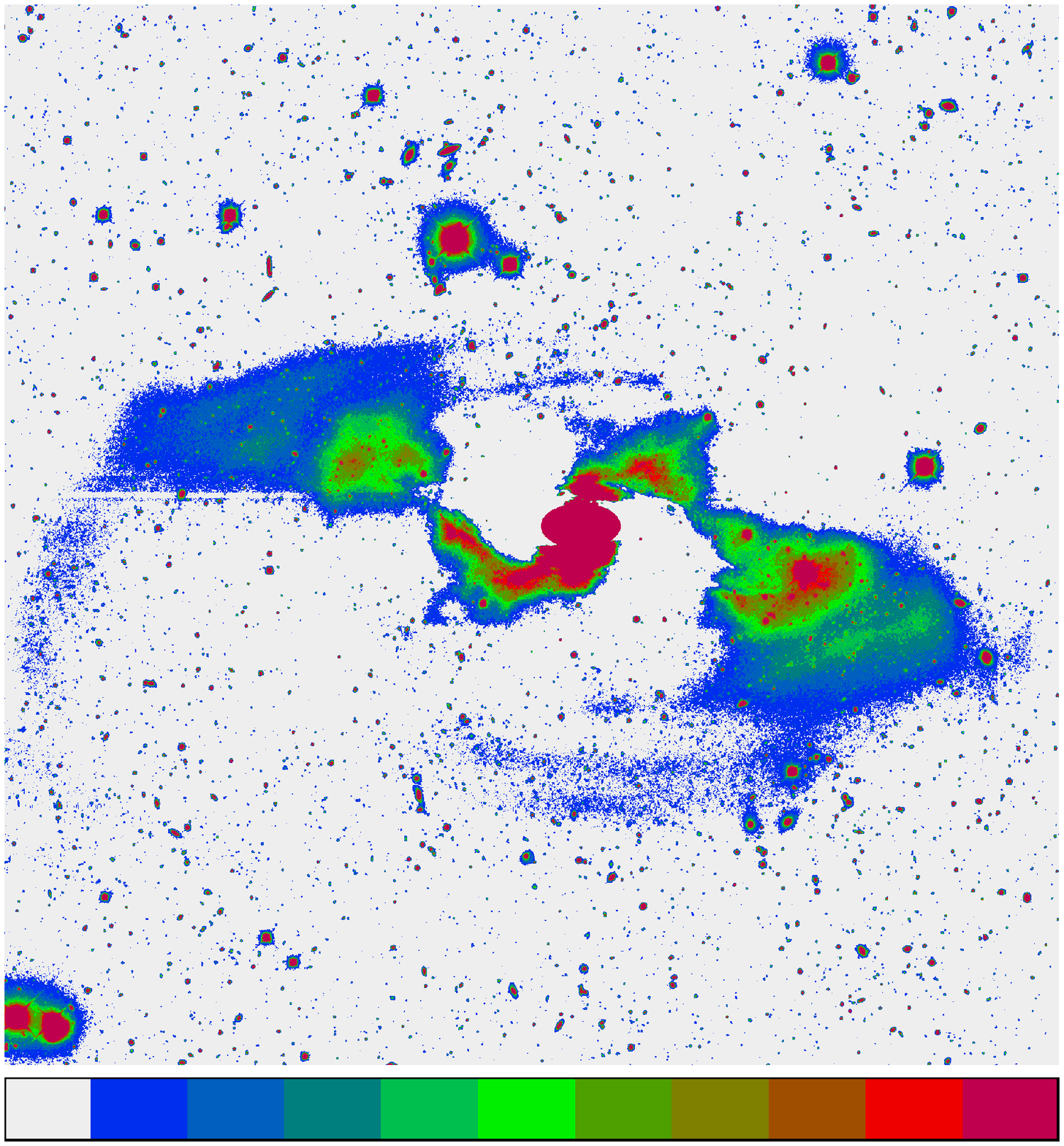}\\
\end{flushright}
\includegraphics[width=85mm]{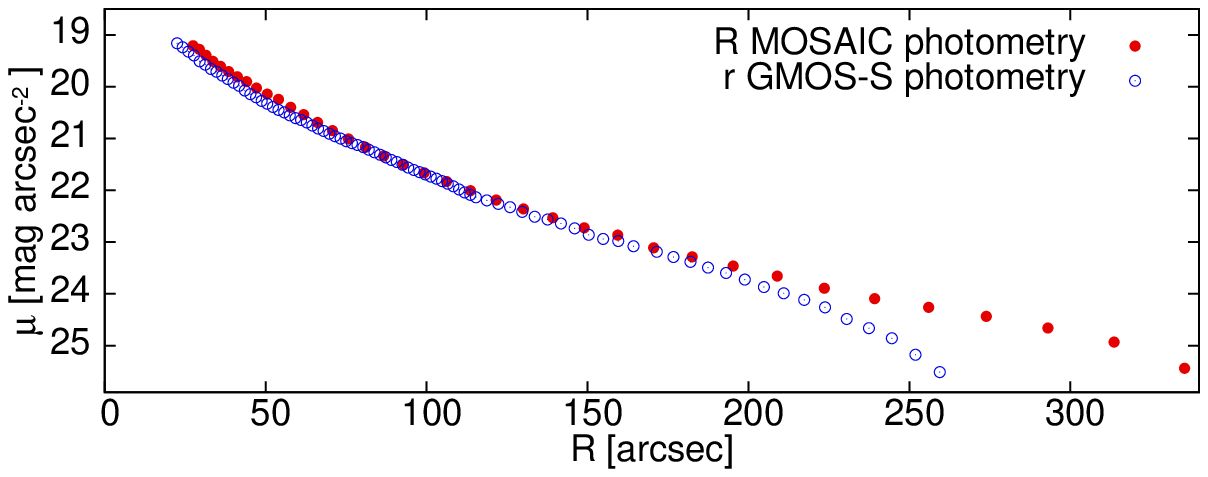}\\
\caption{{\bf Upper panel}: Residuals map, obtained from the 
subtracion of the ELLIPSE galaxy model to the MOSAIC\,II $R$ 
image. The image spans $12 \times 12\,{\rm arcmin}^2$. 
North is up, East to the left. {\bf Lower panel}: Smooth luminosity
profile of NGC\,4753 obtained from the MOSAIC\,II $R$ (filled 
circles), and from the GMOS-S $r'$ (open circles) images. The 
latter one was scaled assuming $r'-R=0.25$ \citep{fuk95}.}    
\label{map}    
\end{figure}

The filled circles in the 
lower panel of Figure\,\ref{map} represent the profile in the 
$R$ filter. For comparison, the surface brightness profile in 
the $r'$ filter was overimposed, assuming $r'-R=0.25$ 
\citep{fuk95}. Both profiles match within 200\,arcsec. The 
differences for larger radii are due to the smaller GMOS field 
of view.
The upper panel from Figure\,\ref{map} shows the residual map
obtained from the subtraction of the $R$ galaxy model to the
original image. The image spans $12 \times 12\,{\rm arcmin}^2$.
 There is clear evidence of underlying substructure. The most 
striking features are two opposite lobes, that seem to be 
connected to the galaxy centre by spiral remmants. Sparser
spiral remmants are identified in the outskirts.

\section{Discussion}

\subsection{The paucity of GCs in galaxies with Type Ia Supernovae}

The work on Supernovae rates, specially over the last years has led to a 
current knowledge of this subject where star formation rates (hereafter SFR) 
are unarguably linked to SN rates \citep{cap99,man05,sul06}.

While it is natural to understand the presence of core-collapse SNe in 
starburst galaxies,  the case of type Ia SNe in early-type galaxies is 
particularly intriguing.

In their study of the correlation between age, metallicity and SNIa peak 
magnitude, \citet{gal08} find that the presence of SNIa in an 
elliptical galaxy is not dependent on its age or metallicity. They claim, 
however, that peak Ia luminosities are correlated with age, while the trend 
observed with metallicity is an artifact.  In addition, they observe that the 
rate of Ia SNe is higher in E/S0 galaxies with spectroscopic ages younger 
than 3 Gyr than in their older counterparts.

Given the mounting evidence favouring the SFR-SNeIa connection, one would 
expect to see young or intermediate-age populations in E/S0 galaxies that 
hosted SNIa. Indeed, some striking examples exist of early-type galaxies 
hosting SNIa's (NGC\,1316 \citealt{str10}, NGC\,4374 \citealt{can11},
NGC\,5128 \citealt{jha07}). 

In NGC\,1316, intermediate-age globular clusters have been spectroscopically 
suspected \citep{gom01} and spectroscopically confirmed \citep{gou01}. Also the 
kinematics of 
this GCS is particularly complex, probably as a consequence of the 2\,Gyr 
merger event \citep{ric14}. Moreover, the GCs colour distribution presents
a distinctive composition, with a bright population in the bulge that
might be dominated by young and intermediate age GCs  \citep{ric12a}.

A common feature of these GCSs, which is observed in NGC\,4753 as well, seems 
to be a low specific frequency \citep[e.g.][]{gom01,ric12a}. This could be 
explained by a dearth of globulars or by a luminosity excess caused, for 
instance, by the presence of younger stellar populations.

\subsection{NGC\,4753 and the general scheme of lenticulars evolution}
In the current scheme, the evolution from spirals to lenticular galaxies 
in dense environments is due to their interaction with the media or with 
their close neighbours \citep{moo99,bek02,sun06,bar07,wil12b,joh14}. However, 
this mechanisms fail to explain the existence of S0s in low density 
environments, like the one where NGC\,4753 is located in, and alternative 
origins should be considered \citep{sto04}. \citet{tal09} studied a sample 
of bright field Es and considered that the tidal features found in the 
diffuse light distribution of most of them, are evidence of mostly recent 
mergers. In that sense, the tidal features in NGC\,4753 could also be related 
to a past minor merger. 

On the basis of a ``mock'' catalogue created from the Millennium Simulation, 
\citet{nie10} have shown that almost all isolated E galaxies have undergone 
merging activity during their formation history. This is supported by the 
evidence of merger events in S0 galaxies in low density environments, e.g. 
the case of NGC\,6861 which was identified by \citet{esc15} by studying 
their GCs. Inhomogeneities in the projected spatial distribution of GCs, 
as spiral-like features, streams and/or arcs, have also been found in other 
early-type galaxies and interpreted as remnants of past mergers 
\citep{dab13,dab14a,dab14b}. On the other side, several galaxies that are 
likely remnants of recent mergers have GCs of intermediate-age 
(about $1 - 3$\,Gyr), as shown by means of combined optical and NIR 
photometry by \citet{tra14}. That is, in different ways, GCs may be part 
of such intermediate-age populations.

In the framework of the Project ATLAS$^{3D}$ \citep{cap11}, a study of the 
distribution and kinematics of CO gas in a sample of early-type galaxies 
\citep{ala13} has been presented, being NGC\,4753 one of their targets. According to the 
CO maps, this galaxy is assigned a morphological classification M (``Mildly 
disrupted'') due to the intrincated dust filaments. Although generally dust 
and CO have similar distributions, NGC\,4753 is one of the clear cases in 
which dust is more extended than the detected CO (their Fig.\,3). The CO 
velocity map of NGC\,4753 is presented in their Fig.\,2 and A20, showing that 
NGC\,4753 is a massive regular rotator. On the basis of the firm statement f
rom \citet{ste92}, regarding that the dust distribution may be understood as 
a precessing disc, the molecular CO disc can be also considered unsettled, 
with an external origin (e.g. an accretion event).

Unusual behaviours in the colour distribution \citep[e.g.][]{gom01,ric12a,cas13b,esc15},
 or GCLF \citep[e.g.][]{gou07} have been considered evidence for GC subpopulations
with intermediate ages, asociated with late mergers. In some cases, it has been
spectroscopically confirmed \citep{gou01,str04b}. These features are also present
in NGC\,4753 GCS, pointing to the presence of GCs with intermediate ages, probably 
related to the same stellar population that produced the SNIa. In this sense, 
another hint in NGC\,4753 is the spatial projected distribution of GCs, which
differs from the usual scenario in early-type galaxies, with red GCs concentrated
towards the galaxy, while blue ones form a more extended halo \citep[e.g.][]{bas06a,ush13,dur14}.

\section{Summary and Conclusions}
In this work, the first detailed optical study of the GCS surrounding the galaxy 
NGC\,4753 is presented. This bright galaxy ($M_V \approx -22.3$), classified as 
lenticular as well as irregular, presents a remarkable dust structure with large 
filaments and is located in a low-density environmet. All these characteristics 
make it a good target to study galaxy evolution in the context of the current galaxy 
formation theories.

On the basis of GMOS $g'$, $r'$, $i'$ images obtained at Gemini-South Observatory, 
and Mosaic\,II $R$ images from the Blanco 4-m telescope at Cerro Tololo Interamerican 
Observatory, we performed the photometry of the GC candidates and surface photometry 
of the galaxy itself. As expected for this galaxy, several properties of the GCS are 
peculiar. The GC colour distribution can be well fitted by a trimodal set of 
blue/intermediate/red subpopulations, instead of just the common blue/red one. The 
fraction of blue GC candidates is similar to that of intermediate plus red ones. The 
intermediate-colour subpopulation can be explained as a SSP of an intermediate age, 
i.e. $1-3$\,Gyr.
The GCLF can be fitted by a Gaussian, but an excess of bright globulars is present 
for $i' > 23$\,mag. If the projected distribution of such a bright subsample is 
analyzed, it shows inhomogeneities in their azimuthal distribution, with fewer 
candidates towards the North. Integrating the GCLF, an estimation of $1070\pm120$\,GC 
candidates is obtained for the whole GCS, which leads to a specific frequency of 
$S_N = 1.3\pm0.15$, also a quite low value for such a massive galaxy.

The surface photometry of the galaxy reveals a complex underlying substructure. In 
addition to the twisted dust lanes and filaments, a pair of inner distorted spiral 
arms linked to another pair of outer ones appear when a smoothed light model is 
subtracted from the $R$ image. Some plums in the outskirts are also visible in the 
residual map. Another peculiarity of NGC\,4753 is that it hosted two SNIa 
(SN1965I and SN1983G), i.e. a high SNIa rate for a lenticular galaxy, that is 
usually related to the existence of intermediate-age population. We suggest that the 
intermediate-colour subpopulation of GCs is probably one of the components of that 
population.

All these properties of NGC\,4753 and of its GCS, converge to a scenario of a quite 
recent accretion event or merger, that might have taken place $1-3$\,Gyr ago. The 
intermediate-age population that might have formed then, may be identified by the 
intermediate colour GC subpopulation and the high SN rate of this galaxy.

\section*{Acknowledgments}
This work was funded with grants from Consejo Nacional de Investigaciones   
Cient\'{\i}ficas y T\'ecnicas de la Rep\'ublica Argentina, Agencia Nacional de   
Promoci\'on Cient\'{\i}fica y Tecnol\'ogica, and Universidad Nacional de La Plata  
(Argentina). MG thanks UNAB/DGID for financial support.\\
Based on observations obtained at 
the Gemini Observatory, which is operated by the Association of Universities 
for Research in Astronomy, Inc., under a cooperative agreement with the NSF on 
behalf of the Gemini partnership: the National Science Foundation (United 
States), the National Research Council (Canada), CONICYT (Chile), the 
Australian Research Council (Australia), Minist\'{e}rio da Ci\^{e}ncia, 
Tecnologia e Inova\c{c}\~{a}o (Brazil) and Ministerio de Ciencia, 
Tecnolog\'{i}a e Innovaci\'{o}n Productiva (Argentina), and on observations
acquired through the Gemini Science Archive.
This research has made use of the NASA/IPAC Extragalactic Database (NED) which 
is operated by the Jet Propulsion Laboratory, California Institute of Technology, 
under contract with the National Aeronautics and Space Administration.
We thank the refereee for his/her suggestions which greatly improved this article.

\bibliographystyle{mn2e}
\bibliography{biblio}

\end{document}